\documentclass[]{mn2e}
\usepackage{graphicx}

\title{A multifrequency study of giant radio sources \\ III. Dynamical age vs. spectral age
of the lobes of selected sources}
\author[J. Machalski et al.]
        {J. Machalski$^1$$\thanks{E-mail: machalsk@oa.uj.edu.pl (JM); jamrozy@oa.uj.edu.pl (MJ); 
       djs@ncra.tifr.res.in (DJS)}$, M. Jamrozy$^1$ and D.J. Saikia$^{2}$ \\ 
$^1$ Obserwatorium Astronomiczne, Uniwersytet Jagiello\'nski, ul. Orla 171, 30244 Krak\'ow, Poland \\
$^2$ National Centre for Radio Astrophysics, TIFR, Pune University Campus, Post Bag 3, Pune 411 007, India }
\date{Accepted.    Received }

\pagerange{\pageref{firstpage}--\pageref{lastpage}}

\begin{document}

\maketitle

\label{firstpage}

\begin{abstract}
The dynamical ages of the opposite lobes of selected giant radio sources are estimated
using the DYNAGE algorithm of Machalski et al., and compared with their spectral
ages estimated and studied by Jamrozy et al. in Paper II. As expected, the DYNAGE fits
give slightly different dynamical ages and other model's parameters for the opposite lobes
modelled {\sl independently} each other, e.g. the age ratios are found between $\sim$1.1
to $\sim$1.4. Demanding similar values of the jet power and the radio core density for
the same source, we look for a {\sl self-consistent} solution for the opposite lobes,
which results in different density profiles along them found by the fit. We also show
that a departure from the equipartition conditions assumed in the model, justified by
X-ray observations of the lobes of some nearby radio galaxies, and a relevant variation
of the magnetic-field strengths may provide an equalisation of the lobes' ages.
A comparison of the dynamical and spectral ages shows that a ratio of the dynamical age
to the spectral age of the lobes of investigated giant radio galaxies is between $\sim$1
and $\sim$5, i.e. is similar to that found for smaller radio galaxies (e.g. Parma et al.
1999). Supplementing possible causes for this effect already discussed in the literature,
like uncertainty of assumed parameters of the model, an influence of a possible departure from
the energy equipartition assumption, etc., the further two are pointed out 
and discussed: (i) a difference between the injection spectral indices describing the
initial energy distributions of the emitting relativistic particles determined using the
DYNAGE algorithm in the dynamical analysis and in the classical spectral-ageing analysis,
and (ii) a different influence of the axial ratio of the lobes in estimation of the dynamical
age and the spectral (synchrotron) age. Arguments are given to suggest that DYNAGE can better
take account of radiative effects  at lower frequencies than the spectral-ageing analysis.
The DYNAGE algorithm is especially effective for sources at high redshifts, for which an
intrinsic spectral curvature is shifted to low frequencies.
\end{abstract}

\section{Introduction}

There are several approaches to estimate the age of a classical double radio
source, beginning with a use of the projected linear size of that source and
estimating the speed of its expansion to that size. Direct measurements of the
expansion speed inferred from the proper motions of the hot spots in compact
symmetric radio sources gave values of about 0.2$c$ -- 0.3$c$ (cf. Owsianik,
Conway \& Polatidis 1998; Owsianik \& Conway 1998). An extrapolation of these 
motions back in time indicates very young ages of such sources, being of the order of
$10^{2}-10^{4}$ years. On the other hand, all the analytical models of the
dynamics and radio-emission properties of powerful double-lobed radio sources (e.g.
Scheuer 1974; Begelman \& Cioffi 1989; Falle 1991; Nath 1995; Kaiser, Dennett-Thorpe
\& Alexander 1997; Blundell, Rawlings \& Willott 1999; Manolakou \& Kirk 2002; 
Kino \& Kawakatu 2005) predict that
those speeds reduce gradually by an order of magnitude or even more with the
source age.

The above has been confirmed by the classical spectral-ageing analysis. There
is no doubt that radio continuum spectra in different parts of an extended
radio source contain important information about the various energy losses and
gains of the radiating particles during the lifetime of the source. With the
assumptions that (i) these particles are immersed in a uniform magnetic
field, (ii) they are not significantly reaccelerated within the source (their
lobes), and (iii) there are no significant mixing of new and old particles --
the observed radio spectrum should steepen with increasing distance from the place
of the last acceleration, i.e. from the hot spots. These predictions have been detected
in many radio sources and used to estimate the radiative ages of the emitting
particles and the expansion speeds in several samples of powerful 3CR sources
(e.g. Myers \& Spangler 1985; Alexander \& Leahy 1987; Leahy, Muxlow \& Stephens 1989;
Carilli et al. 1991; Liu, Pooley \& Riley 1992), in samples of low-luminosity and
medium-luminosity radio galaxies (e.g. Klein et al. 1995; Parma et al. 1999),
as well as in samples and/or of individual `giant'-sized radio sources (e.g.
Lacy et al. 1993; Saripalli et al. 1994; Mack et al. 1998; Schoenmakers et al.
1998, 2000; Lara et al. 2000). 

However, the observed steepening of the spectrum need not be entirely due to
radiative energy losses. A possible evolution of the local magnetic fields,
a bulk backflow and significant mixing of the lobe material, or the difficulties
in disentangling the effects of the various loss processes have been pointed
out in a number of papers (cf. Rudnick, Katz-Stone \& Anderson 1994; Eilek \& Arendt 
1996; Jones, Ryu \& Engel 1999). The spectral steepening due to the energy losses is 
parameterised by a single `break' in the observed spectrum, $\nu_{\rm br}$. The spectral age
of the particle population with the observed $\nu_{\rm br}$ within a constant
magnetic field of strenght $B$ is proportional to $B^{-3/2}\nu_{\rm br}^{-1/2}$.
If $B$ decreases while the source (its lobes) expands, the spectral age should
overestimate the true age of the source. However this age is usually found to be
lower than the source's age inferred from the dynamical considerations (cf.
Kaiser 2000, hereafter referred to as K2000).

For the first time the problem of how to reconcile the spectral and dynamical
ages was undertaken by Blundell \& Rawlings (2000). They discussed how these
two can be quite discrepant from one another rendering use of the classical
spectral ageing method inappropriate. Moving beyond the traditional bulk backflow
picture and considering alternative means of the transport of high-energy particles,
the authors explained the spectral steepening along the lobes not predominantly by
synchrotron ageing but by gentle gradients in the magnetic field. They contended
that spectral ages can give meaningful estimates of dynamical ages only when
these ages are less or much less than $10^{7}$ years. The same problem was
studied by K2000 who extended the
spectral-ageing methods including the underlying source dynamics into the age
estimates. The author claimed that if the bulk backflow and energy losses of the
relativistic electrons, both radiative and adiabatic, are self-consistently taken
into account, the discrepancies between spectral ages and dynamical ages arising
from the earlier methods can be resolved. However analysing the K2000 model,
Machalski et al. (2007) realised that in a majority of the extended FRII-type
radio sources -- even those without distorted lobe structures -- 
the surface-brightness profiles are far from the expected smooth shapes, making
the fitted free parameters of the model highly uncertain. Besides, the K2000
method requires rather high-resolution observations of the radio lobes. But
usually high-resolution observations of low-brightness sources cause a serious
loss of the flux density, so that the K2000 method can in fact only  be applied
to the strongest sources such as Cyg\,A.

In Jamrozy et al. (2008, Paper II of this series), multifrequency observations
with the Very Large Array (VLA) and the Giant Metrewave Radio Telescope (GMRT)
(Konar et al., 2008, Paper I of this series) have been used to determine the
spectral ages of ten selected giant-sized radio
galaxies (GRGs). Using the classical spectral-ageing approach and applying two
different formulae for the equipartition magnetic field estimates: the classical
formula of Miley (1980) and the revised formula of Beck \& Krause (2005) -- the
spectral age distributions along the main axis of the lobes were analysed. In
that paper we found that practically the statistics of the derived age is
independent of the equipartition field formula applied, though those ages can
be quite different for individual sources. We also analysed the injection
spectral indices characterising an initial power-law energy distribution of the
emitting particles. Those indices, determined by fit to the observed radio spectra
with the SYNAGE algorithm of Murgia (1996) were found to be correlated either
with luminosity or redshift, as well as with linear size of the source.

In this paper dynamical ages of the lobes of the ten GRGs studied in Paper II
are determined and compared with their spectral ages. Using the DYNAGE algorithm
(Machalski et al. 2007), we derive the age, the average expansion velocity of
the lobes' heads, the {\sl effective} injection spectral index which approximates
the initial electron continuum averaged over a very broad energy range and over
the present age of source, and other dynamical properties of the sources like
their jets' power, central density near the radio core which determines the local
environment density in which the jets propagate, and the internal pressure in the
lobes. In Section 2 we describe briefly the DYNAGE algorithm for fitting
the dynamical parameters of the model to the observational data given in
Section 3. In Section 4 we present the results of the fit, while the discussion
of the results, especially of the two factors causing a difference between the
dynamical and the spectral ages are given in Section 5. 

\section{The DYNAGE algorithm}

This algorithm is an extension of the analytical model for the evolution of
FRII-type radio sources combining the dynamical model of Kaiser \& Alexander
(1997) with the model for expected radio emission from a source (its lobes or
cocoon) under the influence of energy loss processes published by Kaiser, Dennett-Thorpe
\& Alexander (1997, hereafter KDA model). One of the basic assumptions of the KDA model is
a {\sl continuous} delivery of kinetic energy from the active galactic nucleus (AGN)
to the radio lobes through the jets (e.g. Falle 1991).  The jets terminate in strong
shocks where the jet particles are accelerated and finally inflate the cocoon. The density
distribution of unperturbed external gas is approximated as 

\begin{equation}
\rho(r)=\rho_{0}\left(\frac{r}{a_{0}}\right)^{-\beta};\hspace{5mm}{\rm for}\hspace{5mm} r\geq a_{0},
\end{equation}

\noindent
where $\rho_{0}$ is the central density at the core radius $a_{0}$, and the exponent
$\beta$ describes the density profile in the simplified King's (1972) model. Such a
distribution of the ambient medium is assumed to be invariant with redshift. As in KDA,
the lobe (i.e. half of the cocoon) is approximated by a cylinder of length $D$ and
base diameter $b$, so that its volume $V_{\rm c}$ is determined via the axial ratio,
$R_{\rm T}=D/b$. The cocoon expands along the jet axis driven by the hotspot plasma pressure
$p_{\rm h}$ and in the perpendicular direction by the cocoon pressure $p_{\rm c}$. The
ratio of these pressures is constant for a given radio lobe and depends on its $R_{\rm T}$.
In our calculations  we use the empirical formula taken from K2000

\[{\cal P}_{\rm hc}\equiv p_{\rm h}/p_{\rm c}=(2.14-0.52\beta)R_{\rm T}^{2.04-0.25\beta}.\]

\noindent
Thus the model predicts a self-similar expansion of the cocoon (lobe) and gives analytical
formulae for the time evolution of its geometrical and physical parameters, e.g. the
length of the lobe

\begin{equation}
L(t)=c_{1}\left(\frac{Q_{\rm jet}}{\rho_{0}a_{0}^{\beta}}\right)^{1/(5-\beta)}
t^{3/(5-\beta)},
\end{equation}

\noindent
and the cocoon pressure

\[p_{\rm c}(t)=\frac{18c_{1}^{(2-\beta)}}{(\Gamma_{\rm x}+1)(5-\beta)^{2}{\cal P}_{\rm hc}}
\left( \rho_{0}a_{0}^{\beta}\right)^{3/(5-\beta)}  \]
\begin{equation}
\times Q_{\rm jet}^{(2-\beta)/(5-\beta)}t^{-(4+\beta)/(5-\beta)},
\end{equation}

\noindent
where $c_{1}$ is a dimensionless constant, $\Gamma_{\rm x}$ -- the adiabatic index of
unshocked medium surrounding the cocoon (lobe), $t$ -- the time elapsed since the jet
started from the AGN, i.e. it is the source's (here its lobes') actual age. The cocoon's
pressure determines the energy density within it via the adiabatic index $\Gamma_{\rm c}$
of the cocoon as a whole, depending on the relative pressures of relativistic electrons,
thermal particles, and magnetic `fluid'

\[u_{\rm c}(t)=p_{\rm c}(t)/(\Gamma_{\rm c}-1),\]

\noindent
where $u_{\rm c}=u_{\rm B}+u_{\rm e}(1+k^{\prime})$, with the equipartition condition

\begin{equation}
\xi\equiv\frac{u_{\rm B}}{u_{\rm e}(1+k^{\prime})}=\frac{1+p}{4}=\frac{\alpha_{\rm inj}+1}{2},
\end{equation}

\noindent
where $k^{\prime}$ is the ratio of the energy density of thermal particles to that of
the electrons, and $p$ is the power index of the power-law energy spectrum of radiating
particles. The magnetic field (assumed to be completely tangled) with the energy density
$u_{\rm B}$ and adiabatic index $\Gamma_{\rm B}$ satisfied the relation

\begin{equation}
B(t)\propto u_{\rm B}^{1/2}(t)\propto t^{- \left[\frac{(4+\beta)\Gamma_{\rm B}}
{2(5-\beta)\Gamma_{\rm c}}\right]}.
\end{equation}

\noindent
As in KDA, we assume that the jets consist of an electron--positron plasma and that the
energy contribution from relativistic protons is completely negligible.

The above dynamical
equations are supplemented with an integral giving the total radio emission from the
cocoon at a frequency $\nu$. The cocoon is split into many small volume elements $\delta V$,
each of which is allowed to evolve by expanding adiabatically, i.e. changing the pressure
from the hotspot pressure $p_{\rm h}(t_{\rm i})$ to the cocoon pressure $p_{\rm c}(t_{\rm i})$
as a function of injection time $t_{\rm i}$.
Tracing the effects of adiabatic expansion, synchrotron losses (with the assumed effective
isotropisation of the particles' pitch angle distribution), and inverse Compton scattering
on the cosmic background radiation in the volume elements independently, the radio power
of the cocoon $P_{\nu}$ at a fixed observing frequency is obtained by summing up the
contributions from all elements, resulting in the integral over $t_{\rm i}$ (equation (16)
in KDA). The integral is not analytically solvable and has to be calculated numerically.

The DYNAGE algorithm allows one to determine the values of four of the model's free parameters,
i.e. the jet power, $Q_{\rm jet}$, central core density, $\rho_{0}$, injected spectral index,
$\alpha_{\rm inj}$, and dynamical age, $t$. The determination of their values is possible by
the fit to the observational parameters of a source: its projected linear size, $D$, the
volume of the cocoon, $V_{\rm c}$, the radio luminosity, $P_{\nu}$, and the radio spectrum,
$\alpha_{\nu}$, which provides $P_{\nu_{\rm i}}$ at a number of observing frequencies
$i=1,\,2,\,3\,...$. 

The values of several free parameters of the model have to be assumed. The assumed
values are listed in Table~1, where $\gamma_{\rm i,min}$ and $\gamma_{\rm i,max}$
are the Lorentz factors determining the energy range of the relativistic particles
used in integration of their initial power-law distribution. Since there is a mixture
of three different fluids in the cocoon, this is likely that its equation of state varies
within the cocoon and with time. In the present calculations we choose the KDA `Case 2'
where the cocoon's mixed material has a non-relativistic equation of state but the
energy density of the magnetic field is proportional to that of the relativistic
particles. As discussed in KDA, in this case the value of $\xi$ increases with time
which implies that after some time $u_{\rm B}$ starts to dominate the total energy
density in a given volume element $\delta V$. However at the same time the total volume
of the cocoon increases by a much larger factor, so that the approximation of the
adiabatic index of the whole cocoon of 5/3 is justified. The calculations show that
the model fits obtained either with $\Gamma_{\rm c}$=5/3 or $\Gamma_{\rm c}$=4/3 is
negligible if $k^{\prime}$ is kept constant. More pronounced is influence of a value
of $k^{\prime}$ itself. An increase of $k^{\prime}$ from zero to ten causes a $\sim$10\%
to $\sim$20\% increase of the age due to a thermal expansion of the lobes (for the largest
values of $R_{\rm T}$ it can be more than 30\%, cf. Machalski
et al. 2008). This effect, though ruling the entire time scale,
does not affect age differences between the opposite lobes (as far as $k^{\prime}$ values
are comparable in both lobes). However, without an independent evidence about the fractional
content of thermal particles, holding its value as zero seems to be acceptable when we
analyse the above differences.

Furthermore, we assume the
orientation of the jet axis to the observer's line of sight, $\theta$. Because of
the giant linear sizes of the investigated sources, we assume $\theta=90^{\rm o}$
for almost all the lobes except two  for which the observed asymmetries in
the lobes' separation and/or in their brightness suggest $\theta<90^{\rm o}$.
The details are given in Section 3. An extensive discussion of limitations of the
DYNAGE method and dependence of the age solution on the assumed values of the model's
free parameters are given in Machalski et al. (2007). For example, changing the value
of $\gamma_{\rm i,min}$ from 1 to 10 decreases the age of a source (lobe)  by about
2\,\%--5\,\%, while varying the value of $a_{0}$ between 20 kpc and 5 kpc changes the age
from about $-$10\,\% to about +20\,\%, which is not larger than the age
uncertainties determined for the GRGs analysed in this paper (cf. Table 3).

\begin{table}
\caption{ Assumed values of the model free parameters}
\begin{tabular}{@{}cccccccc}
\hline
$a_{0}$ & $\beta$ & $\gamma_{\rm i,min}$ & $\gamma_{\rm i,max}$ &
$\Gamma_{\rm c}$, $\Gamma_{\rm x}$ & $\Gamma_{\rm B}$ & $k^{\prime}$ & $\theta$\\
10\,kpc & 1.5 & 1 & 10$^{7}$ & 5/3 & 4/3 & 0 & 90$^{\rm o}$\\
\hline
\end{tabular}
\end{table}

\noindent
The fitting procedure consists of three steps:

(1) For a given value of $\alpha_{\rm inj}$ and a number of values of $t$, the values
of $Q_{\rm jet}(\alpha_{\rm inj},t,P_{\nu})$ and $\rho_{0}(\alpha_{\rm inj},t,P_{\nu})$
are determined by the fit of equation (2) to the deprojected linear size of a lobe $D/\sin\theta$
and equation (16) of KDA to the lobe's luminosity at a given frequency $P_{\nu}$.

(2) Performing the above procedure for all of the luminosities chosen to represent the
lobe's spectrum, we get a set of solutions for $Q_{\rm jet}$ and $\rho_{0}$ shown
in Fig.\,1a. As the source (its lobe) must have the same age at any of the observing
frequencies, we search for an age at which the parameters $Q_{\rm jet}$ and $\rho_{0}$
have possibly identical values. A perfect intersection of all the $Q_{\rm jet}-\rho_{0}$
curves corresponding to different observing frequencies at some particular age is
expected if the observed spectrum agrees with the theoretically predicted one in the
framework of the `continuum injection' (C.I.) model of energy losses (cf. Myers \&
Spangler 1985; Carilli et al. 1991). However, the observed spectrum of the analysed
sources (and their lobes) may depart from such a theoretical shape, therefore the
`goodness' of the intersection is quantified by the $\Delta(t)$ measure defined by
equation (4) in Machalski et al. (2007). A minimum of $\Delta(t)$ (i.e. `the best'
intersection of the $Q_{\rm jet}-\rho_{0}$ curves for different frequencies and for
the given value of $\alpha_{\rm inj}$) is considered as an estimate of the source's
(lobe's) dynamical age $t(\alpha_{\rm inj})$ (Fig.\,1b). This minimum also distinguishes
a value of $Q_{\rm jet}(\alpha_{\rm inj})$ (and $\rho_{0}$, $p_{\rm c}$, $u_{\rm c}$,
all dependent on a value of $\alpha_{\rm inj}$).

(3) The steps (1) and (2) are repeated for a  number of $\alpha_{\rm inj}$ values. By
varying this parameter during the fitting procedure, we search for its value which provides
a minimum of the product $Q_{\rm jet}\times t$, i.e. a minimum of the kinetic energy
delivered to the given lobe during the age $t$. Following the original KDA assumption
during integration of the expected radio emission, the minimum energy condition
is initially fulfilled in each volume element of the cocoon, therefore the minimum of
$Q_{\rm jet}\times t$ corresponds to a minimum of the total energy density in it.

\begin{figure*}
\includegraphics[width=17cm]{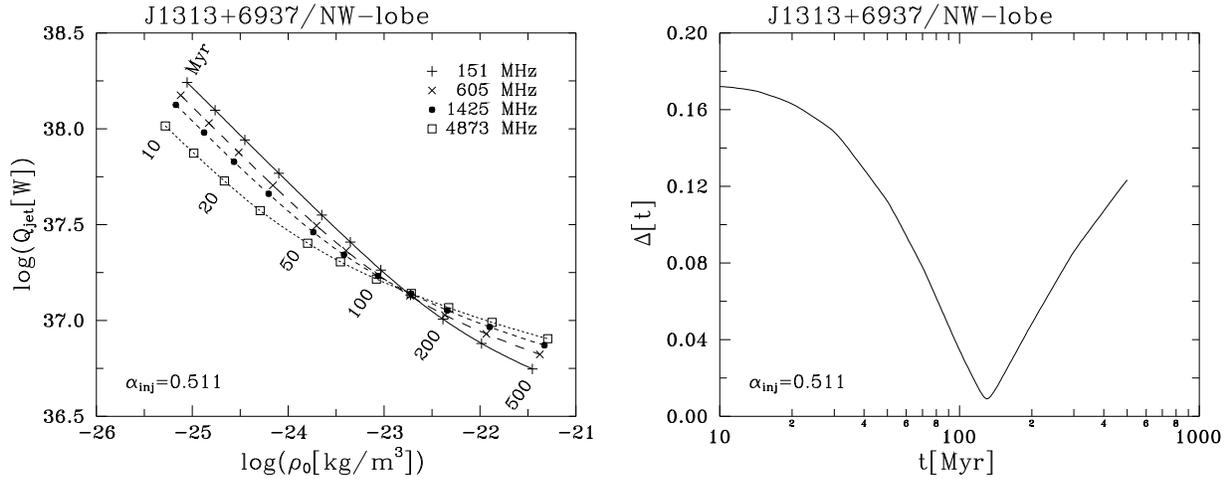}
\caption[]{{\bf a)} $Q_{\rm jet}-\rho_{0}$ diagram for the NW lobe of J1313+6937. The
radio luminosities at the four frequencies indicated and used to fit the model are given
in Table~2. The $\alpha_{\rm inj}$ value for which the jet kinetic energy achieves a
minimum is given. {\bf b)} `Goodness' of fit quantified by the $\Delta$ measure vs. the
age for the same value of $\alpha_{\rm inj}$. Its minimum indicates the age solution
for the given lobe, cf. Table~3.}
\end{figure*}

\section {The data}

The observational parameters of the GRGs under investigation and their lobes are given in
Table~2.  Columns (1), (2), (3), and (7) are self explanatory. Columns (4) and (8) give
linear size $D$ of the opposite lobes (the first line) and their axial ratio $R_{\rm T}$
(the second line); both values with their standard error. The value of $R_{\rm T}$ is
determined on the maps with the highest sensitivity to surface brightness available as the ratio
of $D$ and the largest deconvolved width of transversal cross-sections through the lobe, $b$.
The standard errors of $D$ and $R_{\rm T}$ result from the uncertainties of angular size
of the lobe's length and its base diameter. Columns (5) and (9) give the observing frequencies
selected from the available spectral data and used to calculate the radio luminosities, $P_{\nu}$,
necessary for the fitting procedure (cf. Section 2). For a technical reason, we limited the
fitting procedure to four frequencies only, possibly spanning the largest frequency range. 
The letter preceding the observing frequency indicates the radio telescope (array) or the radio
survey with (or in) which a relevant flux density was measured: G -- for GMRT, V -- for VLA,
and W -- for WSRT. Mostly the flux densities published in Paper I are used, but when necessary,
the spectral data are supplemented with flux densities taken from other observations (surveys):
VLSS (Cohen et al. 2007), 7C (Pooley et al. 1998; Riley et al. 1999), WENSS (Rengelink et al.
1997), B3 (Ficarra et al. 1985), and NVSS (Condon et al. 1998). Where the available radio maps
make it possible, the flux densities of the lobes used to calculate their luminosities are
cleared of the hot spots' emission. Columns (6) and (10) give the derived luminosities
calculated using $H_{0}$=71 km\,s$^{-1}$Mpc$^{-1}$, $\Omega_{\rm m}$=0.27, and $\Omega_{\rm vac}$=0.73.

\begin{table*}
\begin{center}
\caption{Observational parameters of the sources and their lobes}
\begin{tabular*}{172mm}[]{lllrrclrrc}
\hline
IAU name  &$\;\;z$ & Larger &D(kpc) & $\nu\;\;\;$  &log\,P$_{\nu}$ & Shorter  &D(kpc)&  $\nu\;\;\;$ & log\,P$_{\nu}$\\
Other name &     & lobe   &R$_{\rm T}\;\;\;$& (MHz)  &(W\,Hz$^{-1}$sr$^{-1}$) &lobe&R$_{\rm T}\;\;\;$ & (MHz) & 
(W\,Hz$^{-1}$sr$^{-1}$)\\
(1) & (2) & (3) &$(4)\;\;\;$ &$(5)\;\;\;$ & (6) & (7) &$(8)\;\;\;$ &$(9)\;\;\;$ & (10) \\
\hline
J0912+3510 & 0.2489 & S &  901$\pm$19 & 7C151 & 24.881 & N$\dagger$& 578$\pm$82 & 7C151 & 24.556\\
& & &                     5.2$\pm$1.5 &  G606 & 24.436 &           & 4.1$\pm$1.1&  G606 & 24.154\\
                                & & & & V1400 & 24.156 &                      & & V1400 & 23.892\\
                                & & & & V4860 & 23.710 &                      & & V4860 & 23.456\\
J0927+3510 & (0.55) &SE & 1119$\pm$20 & 7C151 & 25.395 & NW & 1087$\pm$32 &       7C151 & 25.253\\
& & &                     4.6$\pm$1.2 &  G606 & 25.005 &    & 4.4$\pm$1.2       &  G606 & 24.788\\
                                & & & & V1400 & 24.711 &                      & & V1400 & 24.486\\
                                & & & & V4860 & 24.236 &                      & & V4860 & 24.000\\
J1155+4029 & (0.53) & SW & 956$\pm$12 &  G241 & 25.346 & NE$\ddagger$&628$\pm$10& 7C151 & 26.343\\
& & &                     8.1$\pm$1.4 &  G605 & 24.968 &           & 3.2$\pm$0.8&  B408 & 25.925\\
                                & & & & V1400 & 24.603 &                      & & V1400 & 25.383\\
                                & & & & V4860 & 24.036 &                      & & V4860 & 24.821\\
J1313+6937 & 0.1064 & SE & 422$\pm$30 & 7C151 & 24.806 & NW &  323$\pm$29 &       7C151 & 24.942\\
DA340      &        &    &3.3$\pm$1.1 &  G605 & 24.389 &    &  2.4$\pm$0.8      &  G605 & 24.516\\
                                & & & & V1425 & 24.095 &                      & & V1425 & 24.229\\
                                & & & & V4873 & 23.593 &                      & & V4873 & 23.759\\
J1343+3758 & 0.2267 & NE &1404$\pm$36 & 7C151 & 24.409 & SW & 1059$\pm$11 &       7C151 & 24.716\\
& & &                     5.4$\pm$1.3 &  W325 & 24.200 &    & 3.3$\pm$0.7       &  W325 & 24.423\\
                                & & & & V1400 & 23.811 &                      & & V1400 & 23.981\\
                                & & & & V4860 & 23.290 &                      & & V4860 & 23.382\\
J1453+3308 & 0.249  & S  & 756$\pm$12 &  G240 & 24.953 & N  & 570$\pm$12  &        G240 & 25.170\\
B1450+333  &        &    &4.8$\pm$1.6 &  G605 & 24.704 &    & 2.8$\pm$1.0       &  G605 & 24.933\\
                                & & & & V1400 & 24.356 &                      & & V1400 & 24.586\\
                                & & & & V4860 & 23.705 &                      & & V4860 & 24.029\\
J1604+3438 & 0.2817 & E  & 443$\pm$17 &  G239 & 24.692 & W  &  403$\pm$21 &        G239 & 24.638\\
& & &                      2.6$\pm$0. &  G614 & 24.361 &    &  2.4$\pm$0.4      &  G614 & 24.300\\
                                & & & & G1265 & 24.198 &                      & & G1265 & 24.107\\
                                & & & & V4860 & 23.603 &                      & & V4860 & 23.526\\
J1604+3731 & 0.814  & NW & 679$\pm$23 &  G334 & 25.708 & SE &  667$\pm$15 &        G334 & 25.804\\
7C1602+376 &        &    &2.4$\pm$0.5 &  G613 & 25.439 &    & 2.8$\pm$0.5       &  G613 & 25.530\\
                                & & & & G1289 & 25.219 &                      & & G1289 & 25.318\\
                                & & & & V4860 & 24.544 &                      & & V4860 & 24.697\\
J1702+4217 & 0.476  & SW & 663$\pm$47 &  W325 & 25.213 & NE &  497$\pm$47 &        W325 & 25.364\\
7C1701+423 &        &    &3.2$\pm$0.7 &  G602 & 25.002 &    &  1.9$\pm$0.5      &  G602 & 25.144\\
                                & & & & V1425 & 24.685 &                      & & V1425 & 24.827\\
                                & & & & V4860 & 24.131 &                      & & V4860 & 24.313\\
J2312+1845 & 0.427  & NE & 544$\pm$34 &   V74 & 26.751 & SW &  512$\pm$28 &         V74 & 26.843\\
3C457      &        &    &4.1$\pm$1.2 &  G334 & 26.177 &    &  3.4$\pm$1.1      &  G334 & 26.284\\
                                & & & & V1425 & 25.615 &                      & & V1425 & 25.738\\
                                & & & & V4866 & 25.087 &                      & & V4866 & 25.213\\
\hline
\end{tabular*}
\end{center}
{\sl Notes.} A letter preceding the observing frequency indicates the radio telescope (array)
used: G -- GMRT, V -- VLA, W -- WSRT, or the surveys: 7C, B3 (cf. the text).
$\dagger$ indicates the lobe with an inclination angle of 70$^{\circ}$
assumed; $\ddagger$ indicates the lobe with an inclination angle of 50$^{\circ}$ assumed. The
corresponding values of R$_{\Theta}$ give the ratio of deprojected length of the lobes. 
\end{table*}

\section{Dynamical age and other physical parameters}

\subsection{Independent solutions for the individual lobes}

In the first step of modelling the age and other physical parameters of the
investigated GRGs, we fit them independently for each of the two lobes of a given
source using their observational data given in Table~2 and the values of other free
parameters of the model from Table~1, in particular with $\beta$=1.5.  
The dependence of the resulting model solutions of $t$, $Q_{\rm jet}$, $\rho_{0}$, 
and $Q_{\rm jet}\times t$ on $\alpha_{\rm inj}$ for the N-lobes of J0912+3510 and
J1155+4029, i.e. for these lobes for which we assumed $\theta<90^{\circ}$, is shown
in Fig.\,2. The vertical line and the arrow indicate the value of $\alpha_{\rm inj}$
that corresponds to the minimum of the kinetic energy delivered to the given lobe by
the jet. The parameter values resulting from the fits are listed in columns (3)--(7)
in Table~3.
 
As might be expected, the DYNAGE solutions of age and other physical parameters for
the opposite lobes of the same source appear somewhat different. The question is
whether these differences are statistically significant. To answer this question, we
estimate an error on the fitted values of a given parameter. The error estimates for the
fitted values of $\alpha_{\rm inj}$ and age ($t$) of the lobes are included in
Table~3. These errors indicate that a difference between the ages of the opposite
lobes is always insignificant; however there is a marginal tendency of the larger
lobes to be older than the opposite shorter ones. It is also worth noting that in
at least three of the ten sources, a difference between the values of the
effective initial spectral index $\alpha_{\rm inj}$ is larger than 2.5--3\,$\sigma$. 
Such differences are possible if an evolution of the magnetic field, various
energy losses and acceleration processes of the relativistic particles, and mixing
of plasma at different ages -- are different at the heads of the opposite lobes.

The above suggests that such differences in the ageing properties of the opposite lobes,
especially in giant-sized radio sources, may be caused by different environmental
conditions. This is likely confirmed by the evident differences in the fitted values
of $\rho_{0}$. Although this parameter denotes a density of the central radio core,
in the DYNAGE algorithm we fit the term $\rho_{0}a_{0}^{\beta}$ which appears in
Eqs.\,(1), (2), and (3).
If the values of $a_{0}$ and $\beta$ are fixed, the fitting procedure gives the
values of $\rho_{0}$ slightly or significantly different for the opposite lobes. However,
we can expect a single value of $\rho_{0}$ (and $Q_{\rm jet}$) for a given source.
Therefore, in the next subsection we describe and check a `self-consistent' solution
for the investigated lobes.

\subsection{Self-consistent solution for the opposite lobes}

In this second step of the modelling we averaged the values of $Q_{\rm jet}$ and
$\rho_{0}$ found from fits for the opposite lobes (given in columns (5) and (6) of
Table~3), and now treat them as the fixed free parameters of the model, 
$\langle Q_{\rm jet}\rangle$ and $\langle\rho_{0}\rangle$, respectively. Given these
values, we can determine a value of $\beta$ for each of the two opposite lobes,
hereafter denoted as $\beta_{\rm s.c.}$.
In order to do that, we assume that a power-law density profile of the ambient
environment can be extended to distances as large as lengths of the lobes of GRGs.
Although such an assumption can be invalid for distances larger than a few hundred
of kpc (cf. Gopal-Krishna \& Wiita 1987; Machalski et al. 2008), the ambient
densities, $\rho_{\rm a}$, calculated for the investigated GRGs from equation (1)
with $r=D$ are still within an acceptable range of density ($\sim 3\times 10^{-28},
\sim 2\times 10^{-25}$ kg\,m$^{-3}$). Thus, equalising $\rho_{\rm a}$ values in the
independent solution and the self-consistent solution, we have from equation (1)

\[\beta_{\rm s.c.}=\frac{{\rm log}(\langle\rho_{0}\rangle / \rho_{\rm a})}
{{\rm log}(D/a_{0})},\]

\noindent
and transforming equation (2) we calculate an expected age of a given lobe (in the
frame of the self-consistent solution) from

\[t_{\rm s.c.}=\left(\frac{D}{c_{1}}\right)^{(5-\beta_{\rm s.c.})/3}
\left(\frac{\langle\rho_{0}\rangle a_{0}^{\beta_{\rm s.c.}}}{\langle Q_{\rm jet}\rangle}
\right)^{1/3}.\]
     
Finally, seeking a preservation of the minimum energy conditions, we search for the
relevant value of $\alpha_{\rm inj}$, hereafter denoted as $\alpha^{(\rm s.c.)}_{\rm inj}$.
The resulting values of $\beta_{\rm s.c.}$, $\alpha^{(\rm s.c.)}_{\rm inj}$,
$t_{\rm s.c.}$, and $(v_{\rm h}/c)_{\rm s.c.}$ which is an average expansion speed of
the lobe derived from the self-consistent solution, are given in columns (8)--(11) of
Table~3.

\begin{figure*}
\includegraphics[width=17cm]{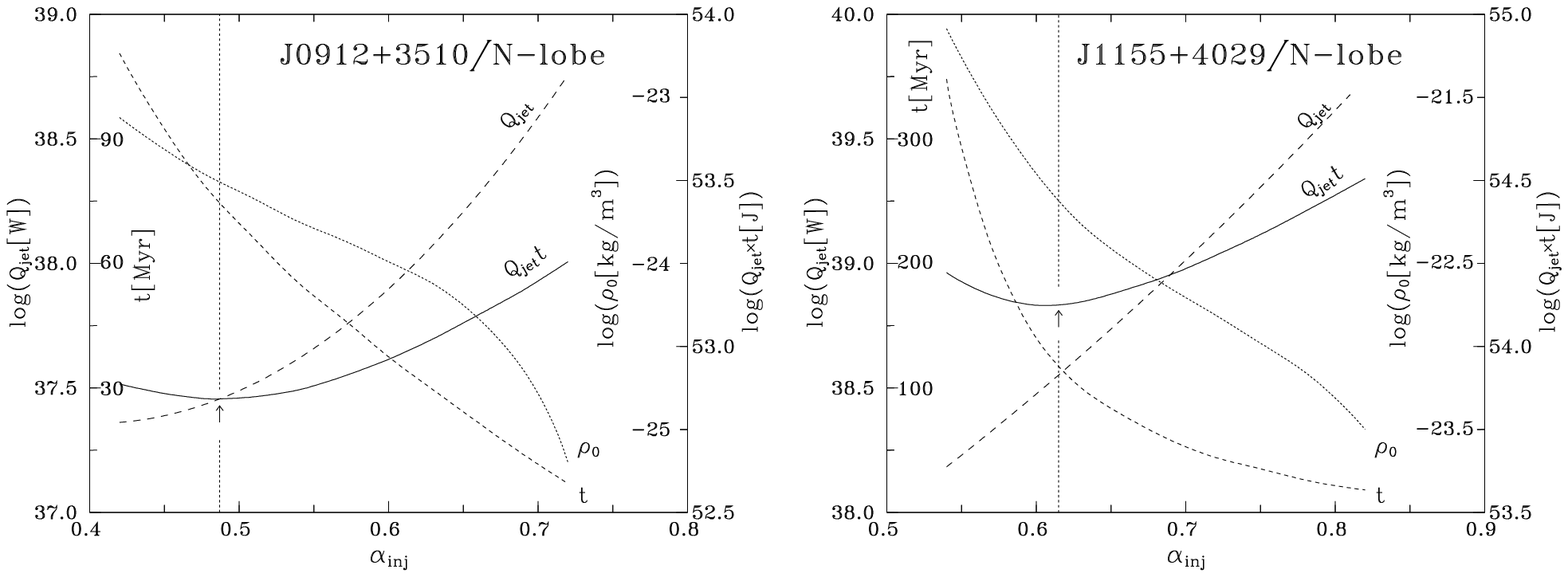}
\caption[]{The dependence of jet power, $Q_{\rm jet}$, age of the lobes, $t$,
central core density, $\rho_{0}$, and kinetic energy delivered to the lobes
during the time $t$, $Q_{\rm jet}\times t$ on the effective initial spectral
index, $\alpha_{\rm inj}$, for the N-lobes of two different GRGs from Table 2.
The minima of the kinetic energy corresponding to the `best solution' of the
age and other physical parameters of the given lobe are marked with the arrows.
The vertical dotted lines indicate proper y-intercepts of each of the four
functions $y=f(\alpha_{\rm inj})$.}
\end{figure*}

\begin{table*}
\begin{center}
\caption{Fitted physical parameters of the lobes. Columns from (3) to (7) give
results of the independent solution, while columns from (8) to (11) give results
of the self-consistent solution}
\begin{tabular*}{162mm}{llcccccccrc}
\hline
Source      & Lobe & $\alpha_{\rm inj}$ & t   & log\,Q$_{\rm jet}$ & log\,$\rho_{0}$ &
v$_{\rm h}$/c & $\beta_{\rm s.c.}$ & $\alpha\rm^{(s.c.)}_{inj}$ & t$_{\rm s.c.}$ & (v$_{\rm h}/c)_{\rm s.c.}$\\
&  &   & (Myr) & (W) & (kg\,m$^{-3}$) &  &  &  & (Myr) \\           
(1) & (2) & (3) & (4) & (5) & (6) & (7) & (8) & (9) & (10) & (11) \\
\hline
J0912+3510 & S  & 0.493$\pm$.006 & 90$\pm$17 & 37.80 & $-$23.30 & 0.033 & 1.45 & $<$0.4& (98) & (0.031)\\
           & N  & 0.487$\pm$.007 & 75$\pm$14 & 37.47 & $-$23.51 & 0.025 & 1.57 & 0.590 &  66  & 0.031 \\
J0927+3510 & SE & 0.490$\pm$.005 & 41$\pm$4  & 38.55 & $-$23.99 & 0.089 & 1.52 & $<$0.4& (44) & (0.083)\\
           & NW & 0.498$\pm$.005 & 50$\pm$5  & 38.38 & $-$23.90 & 0.071 & 1.48 & 0.550 &  46  & 0.077 \\
J1155+4029 & SW & 0.570$\pm$.030 &130$\pm$35 & 38.23 & $-$22.00 & 0.024 & 1.47 & 0.632 & 111  & 0.028 \\
           & NE & 0.615$\pm$.019 &118$\pm$28 & 38.55 & $-$22.13 & 0.017 & 1.54 & 0.569 & 133  & 0.015 \\
J1313+6937 & SE & 0.508$\pm$.011 &152$\pm$22 & 37.18 & $-$22.66 & 0.010 & 1.46 & 0.486 & 153  & 0.010 \\
           & NW & 0.511$\pm$.011 &131$\pm$18 & 37.15 & $-$22.80 & 0.008 & 1.55 & 0.523 & 131  & 0.008 \\
J1343+3758 & NE & 0.461$\pm$.018 & 64$\pm$11 & 37.90 & $-$24.24 & 0.071 & 1.58 & $<$0.4& (69) & (0.066)\\
           & SW & 0.516$\pm$.012 & 94$\pm$13 & 37.76 & $-$23.94 & 0.036 & 1.44 & 0.547 &  87  & 0.039 \\
J1453+3308 & S  & 0.563$\pm$.006 &173$\pm$23 & 37.70 & $-$22.35 & 0.014 & 1.40 & 0.550 & 156  & 0.016 \\
           & N  & 0.518$\pm$.013 &127$\pm$17 & 37.69 & $-$22.92 & 0.015 & 1.71 & 0.544 & 136  & 0.014 \\
J1604+3438 & E  & 0.506$\pm$.007 & 86$\pm$12 & 37.40 & $-$23.40 & 0.017 & 1.50 & 0.440 &  90  & 0.016 \\
           & W  & 0.511$\pm$.008 & 93$\pm$9  & 37.29 & $-$23.39 & 0.014 & 1.50 & 0.547 &  89  & 0.016 \\
J1604+3731 & NW & 0.571$\pm$.008 & 54$\pm$8  & 38.62 & $-$23.48 & 0.042 & 1.45 & 0.590 &  51  & 0.045 \\
           & SE & 0.536$\pm$.009 & 41$\pm$8  & 38.73 & $-$23.70 & 0.054 & 1.57 & 0.512 &  43  & 0.051 \\
J1702+4217 & SW & 0.568$\pm$.012 & 74$\pm$15 & 38.06 & $-$23.07 & 0.030 & 1.38 & 0.544 &  85  & 0.026 \\
           & NE & 0.538$\pm$.012 & 62$\pm$10 & 38.11 & $-$23.73 & 0.027 & 1.76 & 0.544 &  70  & 0.024 \\
J2312+1845 & NE & 0.567$\pm$.012 & 86$\pm$21 & 38.58 & $-$22.10 & 0.022 & 1.47 & 0.574 &  83  & 0.023 \\
           & SW & 0.567$\pm$.013 & 76$\pm$14 & 38.64 & $-$22.23 & 0.022 & 1.54 & 0.561 &  79  & 0.021 \\  
\hline
\end{tabular*}
\end{center}
\end{table*}

\section{Discussion of the results and conclusions}

\subsection{Ages and expansion speeds of the lobes}

The DYNAGE fits imply that (i) the formal ages of the opposite lobes are somewhat different,
though  the difference is not large, given the errors of the fit; (ii) there is a weak trend
in the sense the larger lobe is `older', and (iii) the average expansion speed of the larger
lobe is usually higher than that of the shorter one. The implication (ii)
is consistent with predictions of the simplified kinematic model for the jets' propagation
(e.g. Longair \& Riley 1979): if heads of the jets move through a uniform environment at the
same speed $v_{\rm h}$ at an angle $\theta$ to the line of sight, the shorter lobe will
appear younger by the time delay of $(D_{1}+D_{2}){\rm cot}(\theta)/c$. For the projected
linear size $D_{1}+D_{2}=1$ Mpc and $\theta\approx 70^{\rm o}$, an age difference will be
less than 1.2 Myr. Therefore the derived age differences for the opposite lobes, much
higher than 1--2 Myr, cannot be related to the kinematic effects only, but they likely
indicate actual different jet's propagation conditions in the opposite directions through
the galactic and/or intergalactic medium.
The implication (iii) will be expected if both the lobes are of the same age. Indeed, in spite
of the formal age differences, in seven out of ten GRGs a higher expansion speed is found
for the larger lobe. A faster expansion, in turn, may imply a thinner environment
and/or a higher jet power. Surprisingly, this is not the case for the four sources:
J0912+3510, J1155+4029, J1313+6937, and J1702+4217; the core density fitted for their larger
lobes is higher than that for the shorter ones. Only the shorter (and formally older)
lobes of the sources: J0927+3510, J1343+3758, and J1604+3438 seem to expand slower due to
a denser ambient medium at their side, however the speed differences are insignificant,
being of an order of the speed errors of about 0.005$c$--0.008$c$.

The self-consistent solutions do not change the above picture radically. A median of the age
quotient (older to younger) of about 1.2 is similar in both the independent solutions and in
the self-consistent solutions. Still the expansion speeds tend to be higher for the larger
lobes. Again the exception seems to be  J1604+3731. The values of $\beta_{\rm s.c.}$ found
by the fits are within an acceptable range (1.38, 1.76). It is worth noting that these
values are always less than 2
which is the necessary condition for forming the head of the jet and to observe a source
of the FRII type. On the other hand, the self-consistent solutions require a larger
difference between the effective injection spectral index $\alpha^{(\rm s.c.)}_{\rm inj}$
for the opposite lobes, than those found in the frame of the independent solutions.
The question whether it really reflects different physical conditions governing
the initial energy distribution of the relativistic particles at the head of the lobes,
or it is mostly related to an uncertainty of the fit and/or wrong assumptions in the model, 
is open.

Another alternative self-consistent solution is plausible in which the age
of the opposite lobes has to be the same, and any differences between their size
and luminosity are due to an inhomogeneity (asymmetry) in density distribution of the
surrounding gaseous environment. Again, there is rather no physical circumstances
for significant differences between the jet power and the central core density in
the opposite directions along the jets' axis. In such a scenario, either the same values
of $\rho_{0}$, $Q_{\rm jet}$, and $t$ would be assumed for both lobes or only
postulated, while the values of $a_{0}$, $\beta$, and $\alpha_{\rm inj}$, different for
the two lobes, are determined by the fit. Rearrangement of equation (2) and substitution
of $D$ for $L(t)$ gives

\[t=\left(\frac{D}{c_{1}}\right)^{(5-\beta)/3}\left(\frac{\rho_{0}a_{0}^{\beta}}
{Q_{\rm jet}}\right)^{1/3}.\]

\noindent
Applying this equation separately for either lobe of a source and demanding equality of
their age, jet power, core density, and core radius, we have another equation involving
two unknown quantities: $\beta_{1}$, and $\beta_{2}$, where the fraction $\rho_{0}/Q_{\rm jet}$
is eliminated

\[a_{0}^{(\beta_{1}-\beta_{2})}=
\frac{(D_{2}/c_{1,2})^{(5-\beta_{2})}}{(D_{1}/c_{1,1})^{(5-\beta_{1})}}.\]

\noindent
The values of two latter parameters depend on the energy distributions of particles
injected into the opposite lobes and described by $\alpha_{\rm inj}$ parameters and
on their observed luminosities. This dependence
cannot be expressed analytically, as explained in Section~2. Besides, the values of
$\rho_{0}$ and $Q_{\rm jet}$ (as well as $\alpha_{\rm inj}$) cannot be determined
independently of the unknown values of $a_{0}$ and $\beta$. Therefore, such an
`alternative self-consistent solution' obviously cannot give any explicit results. 

In order to prove this we attempted to find such a solution for the lobes of
J0912+3510, the sample source with a rare asymmetry where the larger lobe is also
much more luminous than the shorter one. Assuming for both lobes $t=82$ Myr, which is
the average of the values determined in the `independent solution' (cf. Table~3), we
fit the values of $Q_{\rm jet}$ and $\rho_{0}$ for a number of combinations of the model
parameters $a_{0}$, $\beta$ and $\alpha_{\rm inj}$. The results are shown in Fig.\,3,
where the abscissa axis gives the core density $\rho_{0}$ scaled to the core radius
$a_{0}$=10 kpc. Exploring the parameter space we notice that (i) none of 
physically acceptable values
of these parameters can provide comparable values of $Q_{\rm jet}$ for the opposite lobes
(cf. the similar paper of Brocksopp et al. 2007 and their Fig.~4),
and (ii) allowing different values of $Q_{\rm jet}$ and $\rho_{0}$ for either lobe,
the assumed age value can be achieved with {\sl any combination} of $a_{0}$, $\beta$ and
$\alpha_{\rm inj}$ parameters. The above calculations confirm that there is no unique
`alternative self-consistent solution'.

There are two possible causes of different `dynamical age' solutions of the
opposite lobes of a FRII-type source, especially a GRG in which the effect of projection
is likely to be neglected. The first cause, the most probable one, could be related to an
unknown actual distribution of magnetic fields, which may depart from the equipartition
conditions in a different way in either lobe. This is confirmed by X-ray imaging observations
which allow measurements of electron energies of radio lobes and magnetic
fields. The derived electron energy densities often exceed those of  magnetic fields
by a factor of a few to several dozen. For example, the ratio $\xi$ of about 0.16 has been
found in the lobes of the galaxy Cen\,B (Tashiro et al. 1998) and of 0.20 and 0.13 in the
E lobe and W lobe of the galaxy For\,A (NGC\,1316), respectively (Isobe et al. 2006;
Tashiro et al. 2008). Our calculations show that increasing or decreasing the
magnetic-field energy density, $u_{\rm B}$, in respect to the equipartition condition
(given in equation (4)) results in an increase or decrease of the fitted age, respectively, which
agrees with the finding of Parma et al. (1999). For example, the four-fold increase
of the ratio $\xi$ in the N-lobe and the corresponding decrease in the S-lobe of J0912+3510
results in about 10\%--15\% increase or decrease of their fitted age, respectively, i.e.
eliminates the age difference found with the `independent solution'. This is
worth to emphasize that this variation of $\xi$ is not equivalent of changing $\alpha_{\rm inj}$.
In both lobes a minimum of $Q_{\rm jet}\times t$ is still around 0.49, which strongly
supports the conclusion that the minimum of the jets' kinetic energy appearing in the age
solutions is not an artefact of numerical calculations, but has a real physical meaning.
The corresponding {\sl effective} spectral index $\alpha_{\rm inj}$ is always within the
narrow range of $p\approx$ 2.0--2.4 suggested by the `non-relativistic shock' paradigm
for the origin of non-thermal electrons within the heads of FRII-type radio sources
(e.g. Blandford \& Eichler 1987; Heavens \& Meisenheimer 1987).

The second cause, less probable one, could be related to a multi-episod jet activity,
strongly suggested by observations of double-double radio galaxies (DDRGs, e.g. Kaiser
et al. 2000; Saripalli et al. 2002; Brocksopp et al. 2007). If the jet activity has been
interrupted and its multiple episodes happened in opposite directions by chance, an
`effective' age of the lobes would be different. There is, at least, one piece of observational
evidence for such a behaviour, i.e. the northern middle lobe of the galaxy Cen\,A which has
no counterpart in its southern lobe (Morganti et al. 1999). Its spectral age of about 20--30 Myr,
estimated by Hardcastle et al. (2008), is much higher than the time scale of about $10^{3}$ yr
for instabilities in the accretion processes.  

\begin{figure}
\includegraphics[width=8cm]{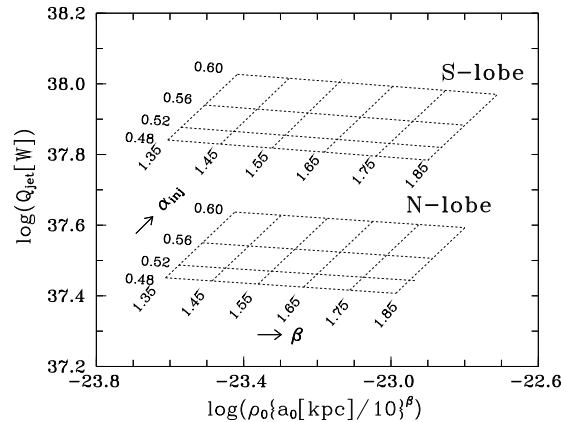}
\caption[]{Model solutions for the lobes of J0912+3510 at the age of 82 Myr
for both lobes. Abscissa gives the (log) core density scaled to the core radius of 10 kpc.
The areas confined by dotted lines indicate very large parameter space
within which both lobes would have the assumed age.}
\end{figure}

\subsection{A relation between the spectral age and the dynamical age}

The spectral age distribution along the jets' axes of the ten investigated GRGs was
analysed in Paper~II. In this subsection we compare (i) the initial spectral indices
resulting from the DYNAGE and SYNAGE fits, and (ii) the dynamical ages derived in this
paper and the synchrotron ages analysed in Paper~II and supplemented in this paper.  As
described in the Introduction,
the notion of a spectral age of the source (or its lobes) is not explicit; it is usually
considered as an age of the emitting particles which can differ in different parts of
the source. Therefore in Paper~II we derived two estimates of this age: the age of the
oldest detected particles in the regions of the observed emission from the lobes, and
the age values resulting from extrapolation of a linear regression of the age on the
distance (i.e. a characteristic speed, $v_{\rm sep}$, which is an indication of the
speed of the lobe material relative to the hot spots) to the radio core. Besides, both
these estimates were derived using two different equipartition magnetic field formulae, 
the classical formula of Miley (1980) and the revised formula of Beck \& Krause (2005).
In each case the spectral age was computed using a spectral break frequency, $\nu_{\rm br}$,
resulting from the SYNAGE fit of the Jaffe \& Perola (1973; JP) model to the radio spectrum
of a given lobe for which the JP fit was mostly the best compared with those
of the Kardashev-Pacholczyk (Pacholczyk 1970; KP) or `continuous injection' (Kardashev
1962; CI) models.

The DYNAGE method for estimating the dynamical age of FRII-type radio sources is based
on the analytical KDA model which assumes a continuous delivery of energy to the lobes
through the jets with a constant power. Therefore, in order to compare the initial
spectral indices resulting from the DYNAGE and SYNAGE fits, we should take into account
the SYNAGE indices fitted with the CI model. However, this model of energy losses does
not account for an adiabatic expansion of the cocoon, while KDA and DYNAGE do.
Therefore, a model much closer to DYNAGE is the CIE model of Murgia (1996) which accounts
for an adiabatic expansion of the volume of source (lobe, cocoon) and a related evolution
of the magnetic field strength in the form $r(t)\propto t^{k}$ ($V(t)\propto t^{3k}$) and
$B(t)\propto t^{m}$. Though each SYNAGE fit of the CIE model to the data points in the
radio spectrum of a given lobe was  the worst as compared with those of the JP, KP, and
CI models -- the resulting spectral ages, $\tau_{\rm CIE}$, should be be closer to
the dynamical ages estimated in this paper. 

The break frequency for the CIE model is $\nu_{\rm br,CIE}=(k-2m-1)^{2}\nu_{\rm br,CI}$.
In the SYNAGE algorithm, where k=1 and m=$-$2, $\nu_{\rm br,CIE}=16\,\nu_{\rm br,CI}$  are
expected for a spectrum of same initial energy distribution, the same age, and final
magnetic field. However, for 20 lobes of the analysed GRGs the ratio $\nu_{\rm br,CIE}/\nu_{\rm br,CI}$
is consistently much less than 16, in the four sample sources is even less than 1. This is caused
by a frequent large difference between $\alpha_{\rm CI}$ and $\alpha_{\rm CIE}$ fitted as a free
parameter in SYNAGE. The values of low-frequency spectral index and break 
frequency for the investigated lobes,
fitted with SYNAGE in the frame of CI and CIE models, as well as the values of corresponding
synchrotron ages are given in Table 4. The value of magnetic field strength $B_{\rm DYN}$,
derived from the magnetic energy density in this paper (cf. equations (4) and (5)) and used to
calculate the ages, is given in column (3). The values of $\alpha_{\rm CI}$, $\nu_{\rm br,CI}$,
$\alpha_{\rm CIE}$, and $\nu_{\rm br,CIE}$  are given in columns (4), (5), (7), and (8),
respectively. The relevant ages, $\tau_{\rm CI}$ and $\tau_{\rm CIE}$, given in columns (6)
and (9), respectively, are calculated from

\begin{equation}
\tau_{\rm CI}=50.3\frac{B^{1/2}}{B^{2}+B^{2}_{\rm iC}}
\{\nu_{\rm br,CI}(1+z)\}^{-1/2}\hspace{1mm}{\rm [Myr]},
\end{equation}

\noindent
and

\begin{equation}
\tau_{\rm CIE}=201.2\frac{B^{1/2}}{B^{2}+B^{2}_{\rm iC}}
\{\nu_{\rm br,CIE}(1+z)\}^{-1/2}\hspace{1mm}{\rm [Myr]},
\end{equation}

\noindent
where $B_{\rm iC}$[nT]=0.318$(1+z)^{2}$ is the inverse-Compton magnetic field strength
equivalent to the cosmic microwave background radiation.  Finally, the differences between
the fitted low-frequency slope and the injected spectral index determined with DYNAGE,
$\alpha_{\rm CI}-\alpha_{\rm inj}$, and $\alpha_{\rm CIE}-\alpha_{\rm inj}$ 
are given in columns (10) and (11), while the ratios between the dynamical age $t$ derive in
this paper and the synchrotron ages $\tau_{\rm CI}$ and $\tau_{\rm CIE}$ -- in columns (12)
and (13), respectively.

\begin{table*}
\begin{center}
\caption{Dynamical and synchrotron ages of the lobes} 
\begin{tabular*}{170mm}{llccccccccccc}
\hline
Source       & Lobe  & $B_{\rm DYN}$ & $\alpha_{\rm CI}$ & $\nu_{\rm br,CI}$ & $\tau_{\rm CI}$
& $\alpha_{\rm CIE}$ & $\nu_{\rm br,CIE}$ & $\tau_{\rm CIE}$ &
$\alpha_{\rm CI}-$ & $\alpha_{\rm CIE}-$ & $t$/  & $t$/ \\
&  & (nT) &  & (GHz) & (Myr) &  & (GHz) & (Myr) &
$\alpha_{\rm inj}$ & $\alpha_{\rm inj}$ & $\tau_{\rm CI}$ & $\tau_{\rm CIE}$  \\
(1) & (2) & (3) & (4) & (5) & (6) & (7) & (8) & (9) & (10) & (11) & (12) & (13)\\
\hline
J0912+3510 & S  & 0.16 & 0.717 &     18.87 & 15.3 & 0.528 & 4.97 & 119. &$\;\;\;$0.224 &$\;\;\;$0.010 & 5.88 & 0.76 \\
           & N  & 0.16 & 0.637 &$\;\;$8.21 & 23.1 & 0.503 & 4.91 & 120. &$\;\;\;$0.150 &$\;\;\;$0.016 & 3.24 & 0.63 \\
J0927+3510 & SE & 0.18 & 0.684 &$\;\;$3.74 & 14.4 & 0.541 & 1.63 & 87.2 &$\;\;\;$0.194 &$\;\;\;$0.051 & 2.85 & 0.47 \\
           & NW & 0.14 & 0.434 &$\;\;$0.30 & 45.7 & 0.554 & 1.77 & 75.3 &$-$0.064 &$\;\;\;$0.056 & 1.09 & 0.66 \\
J1155+4029 & SW & 0.34 & 0.914 &     15.36 & 15.3 & 0.758 & 4.29 & 68.4 &$\;\;\;$0.344 &$\;\;\;$0.188 & 8.47 & 1.90 \\
           & NE & 0.62 & 0.537 &$\;\;$0.11 & 103. & 0.580 &(0.046) &(636)&$-$0.078&$-$0.035 & 1.15 & 0.19 \\
J1313+6937 & SE & 0.23 & 0.613 &$\;\;$2.54 & 70.4 & 0.490 & 0.82 & 496. &$\;\;\;$0.105 &$-$0.018 & 2.16 & 0.31 \\
           & NW & 0.28 & 0.643 &$\;\;$4.29 & 116. & 0.477 & 0.72 & 519. &$\;\;\;$0.132 &$-$0.034 & 1.13 & 0.25 \\
J1343+3758 & NE & 0.08 & 0.571 &$\;\;$2.31 & 35.9 & 0.502 & 1.59 & 173. &$\;\;\;$0.110 &$\;\;\;$0.041 & 1.78 & 0.37 \\
           & SW & 0.11 & 0.646 &$\;\;$1.51 & 50.8 & 0.681 & 6.99 & 94.5 &$\;\;\;$0.130 &$\;\;\;$0.165 & 1.85 & 1.00 \\
J1453+3308 & S  & 0.25 & 0.611 &$\;\;$0.94 & 75.2 & 0.656 & 4.25 & 141. &$\;\;\;$0.048 &$\;\;\;$0.093 & 2.30 & 1.23 \\
           & N  & 0.26 & 0.544 &$\;\;$0.71 & 86.8 & 0.592 & 2.47 & 186. &$\;\;\;$0.026 &$\;\;\;$0.074 & 1.46 & 0.68 \\
J1604+3438 & E  & 0.21 & 0.582 &$\;\;$1.39 & 54.5 & 0.541 & 1.89 & 187. &$\;\;\;$0.076 &$\;\;\;$0.035 & 1.58 & 0.46 \\
           & W  & 0.21 & 0.584 &$\;\;$1.42 & 53.9 & 0.528 & 1.50 & 210. &$\;\;\;$0.073 &$\;\;\;$0.017 & 1.72 & 0.44 \\
J1604+3731 & NW & 0.35 & 0.769 &$\;\;$1.59 & 14.4 & 0.783 & 3.27 & 40.1 &$\;\;\;$0.198 &$\;\;\;$0.212 & 3.75 & 1.35 \\
           & SE & 0.35 & 0.702 &$\;\;$1.48 & 14.9 & 0.732 & 5.13 & 32.0 &$\;\;\;$0.166 &$\;\;\;$0.196 & 2.75 & 1.28 \\
J1702+4217 & SW & 0.32 & 0.665 &$\;\;$1.49 & 32.9 & 0.668 & 4.97 & 72.2 &$\;\;\;$0.097 &$\;\;\;$0.100 & 2.25 & 1.03 \\
           & NE & 0.27 & 0.668 &$\;\;$2.14 & 26.6 & 0.617 & 3.99 & 77.9 &$\;\;\;$0.130 &$\;\;\;$0.079 & 2.33 & 0.80 \\
J2312+1845 & NE & 0.71 & 0.783 &$\;\;$4.08 & 19.0 & 0.656 & 3.27 & 85.0 &$\;\;\;$0.216 &$\;\;\;$0.089 & 4.52 & 1.01 \\
           & SW & 0.76 & 0.770 &$\;\;$4.06 & 18.3 & 0.617 & 2.29 & 97.3 &$\;\;\;$0.203 &$\;\;\;$0.050 & 4.16 & 0.78 \\
\hline  
\end{tabular*}
\end{center}
\end{table*}

\begin{figure*}
\includegraphics[width=17cm]{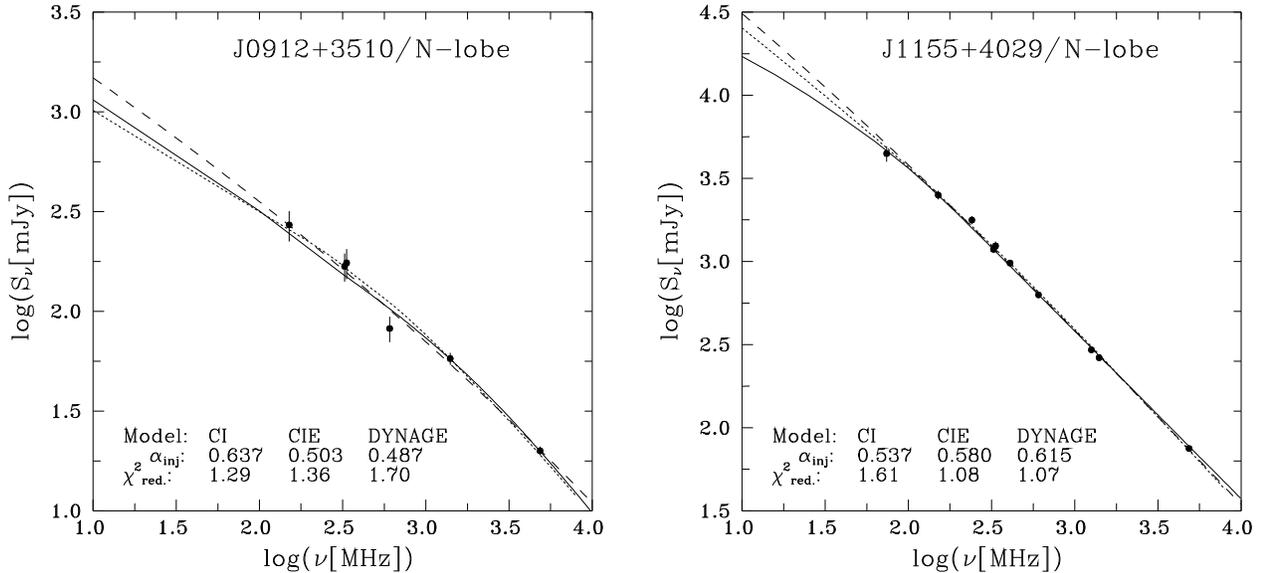}
\caption[]{Exemplary radio spectra of the lobes and fits of the three models of energy losses:
CI (a pure continuous injection) -- the solid line, original CIE (continuous injection and
expansion -- the dashed line, and DYNAGE -- the dotted line. The values of $\alpha_{\rm inj}$
fitted for each of the models and the corresponding reduced $\chi^{2}$ values are given for a
comparison.}
\end{figure*}

The data in Table~4 show that the equipartition magnetic fields derived with the DYNAGE
algorithm are usually stronger than those calculated with the classical Miley's formula.
On the second hand, all these fields are weaker that those calculated with the revised
formula of Beck \& Krause (2005) published in Paper~I. The reason for the latter effect is
that the observed curved radio spectra were integrated, while Beck \& Krause assumed that
the proton spectrum is straight and any steepening in the radio spectrum is due to energy
losses of the electrons. If, however, the proton spectrum is similarly curved as the electron
spectrum, the Beck \& Krause values would be too high. Nevertheless, the resulting spectral
ages are not too much sensitive to the magnetic field strengths, depending
more on the spectral break $\nu_{\rm br}$ and on the ratio $B/B_{\rm iC}$. The examples of
the spectral best fit for the lobes of two different sample sources are shown in Fig.\,4.
Three different fit results for the N lobes of the sources J0912+3510 and J1155+4029  are
compared. Note that the quality of the fits, expressed by the reduced $\chi^{2}$ values,
are comparable within the frequency range for which flux-density data have been available.
The fits clearly show that the frequency range crucial for a better discrimination
between the models is below 100 MHz, i.e. the range where near-future observations with the
LOFAR array will be decisive. 

The data in Table~4 also exhibit unexpected discrepances between the fitted values of 
$\alpha_{\rm CI}$ (or $\alpha_{\rm CIE}$) and  
$\nu_{\rm br}$ in opposite lobes of some sample sources, e.g. J1155+4029 and J1343+3758.
The most discrepant are the values of $\nu_{\rm br,CIE}$ in J1155+4029. This is shown in
Fig.\,5 where the observed spectra of its opposite lobes are compared. Both slopes and
spectral shapes look very similar, nevertheless the formal fitted values of $\alpha_{\rm inj}$
and especially of $\nu_{\rm br,CIE}$ are dramatically different. How much these values are
uncertain is indicated by their errors provided by SYNAGE: 32 MHz$\,<\nu_{\rm br,CIE}<$241 GHz
and 0.08 MHz$<\,\nu_{\rm br,CIE}<$1.18 GHz for the SW lobe and NE lobe, respectively. This is
worth to notice that J1155+4029 has the largest asymmetry in the lobes' luminosities.

\begin{figure}
\includegraphics[width=8cm]{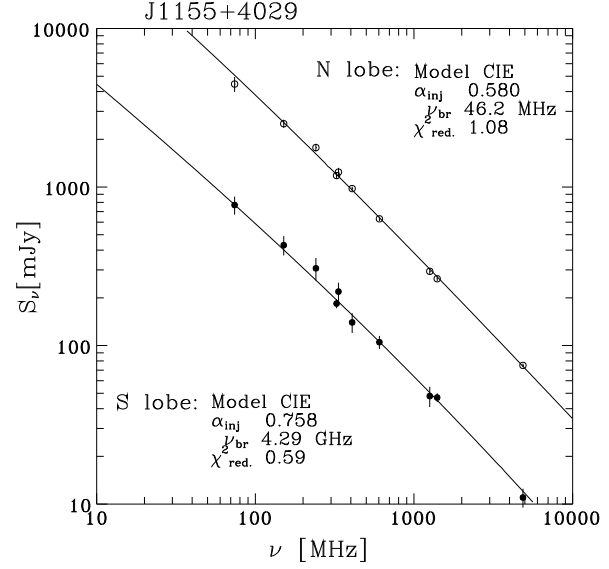}
\caption[]{Radio spectra of the opposite lobes of J1155+4029 and their spectral curvature
parameters fitted with the original CIE model.}
\end{figure}

The fit results indicate that a difference between the dynamical age, $t$, and the
synchrotron age, $\tau_{\rm CI}$ or $\tau_{\rm CIE}$ is related to a difference between
the values of $\alpha_{\rm inj}$ (column (3) of Table~3), $\alpha_{\rm CI}$ or
$\alpha_{\rm CIE}$ (columns 4 and 7 of Table~4). As this was noted in paper of Machalski
et al. (2007), $\alpha_{\rm inj}$ values derived with DYNAGE are usually lower than
those  fitted with SYNAGE. An opposite situation is rare; an example is the NE lobe of
J1155+4029 for which the SYNAGE fit gave $\alpha_{\rm CI}$=0.537$\pm$0.021 and
$\alpha_{\rm CIE}$=0.580$^{+0.13}_{-0.05}$  while $\alpha_{\rm inj}$=0.615$\pm$0.019 is found
in this paper. Such a case is likely caused by a very uncertain observational data at low
frequencies. In general, an increase of the difference $\alpha_{\rm syn}-\alpha_{\rm inj}$
corresponds to an increase of the ratio $t/\tau_{\rm syn}$, where the label `syn' indicates
either $\alpha_{\rm CI}$ or $\alpha_{\rm CIE}$. This correlation (shown in Fig.\,6) is strong;
the correlation coefficient between the above two variables for our twenty lobes is +0.837 in
the case of $t/\tau_{\rm CI}$ and +0.884 in the case of $t/\tau_{\rm CIE}$. The data in
Table~4 show that in 13 of 20 investigated lobes the synchrotron age calculated with $k$=1 and
$m$=$-$2 is higher than the dynamical age determined in this paper (the ratio $t/\tau_{\rm CIE}$
is less than unity in Fig.\,6). Indeed, the corresponding factors in the KDA model of the
jet's dynamics, given by equations (2) and (5), are $k$=6/7 and $m$=$-$11/14 for the assumed
values of $\beta$, $\Gamma_{\rm B}$, and  $\Gamma_{\rm c}$. As a result, we rather would expect
$\tau_{\rm CIE}=(10/7)\,\tau_{\rm CI}$ instead of $\tau_{\rm CIE}=4\,\tau_{\rm CI}$. In fact,
in Fig.\,6 the y-intercepts of $t/\tau_{\rm CI}$ and $t/\tau_{\rm CIE}$ at $\alpha_{\rm syn}-
\alpha_{\rm inj}$=0 differ by $\sim 2$, not by 4. Therefore, we argue that the CI model
underestimates the lobe's age since it does not take into account the expansion losses, while
the CIE one overestimates the lobe's age since the original expansion factors ($k$=1 and
$m$=$-$2) are likely too extreme. In between of these, the DYNAGE age solution should be
better since the expansion parameters are connected to actual geometry of the lobes for each
specific GRGs.

The plot in Fig.\,6 shows that the dynamical age of the GRG's lobes is about 1 to 5 times larger
than their radiative age. Very similar age ratios were found by Parma et al. (1999) for much
smaller, low luminosity radio galaxies. Beside the causes of this effect already discussed in
the literature, there are two other possible explanation of this effect:

(1) As expected, both the synchrotron and the dynamical ages are sensitive to the 
$\alpha_{\rm inj}$ parameter. However, it seems that an $\alpha_{\rm inj}$ value
determined with the DYNAGE method from the minimum kinetic energy of the jets is less
sensitive to an uncertainty of the low-frequency spectrum of a source than values of
$\alpha_{\rm CI}$ or $\alpha_{\rm CIE}$ found with SYNAGE. The only difference between
the spectra provided with the model accounting for the adiabatic losses and the model
not accounting for them -- is a larger range of transition of the spectral slope from
$\alpha_{\rm inj}$ to $\alpha_{\rm inj}+0.5$ in the former model. Therefore, in further
discussion given below, we do not differentiate between $\alpha_{\rm CI}$ and
$\alpha_{\rm CIE}$. The $\alpha_{\rm inj}$ and $\alpha_{\rm CI}$
parameters both determine the slope of the spectrum at frequencies where radiative
losses are not yet important. 

\begin{figure}
\includegraphics[width=8cm]{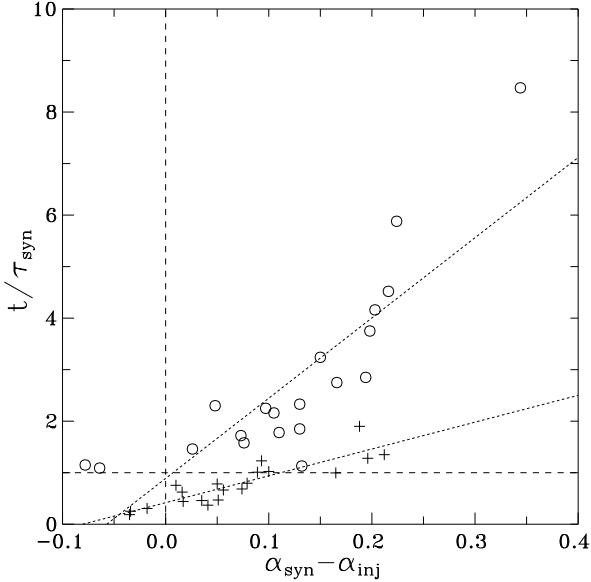}
\caption[]{Ratio of the dynamical age of the lobes, $t$, and their synchrotron age,
$\tau_{\rm syn}$, vs. the difference between $\alpha_{\rm syn}-\alpha_{\rm inj}$ where
$\alpha_{\rm syn}$ means either $\alpha_{\rm CI}$ or $\alpha_{\rm CIE}$.
The circles indicate the ratio of $t$ and $\tau_{\rm CI}$ while the crosses
-- the ratio of $t$ and $\tau_{\rm CIE}$. The two dotted lines show linear regression
lines of the above ratios on the $\alpha_{\rm syn}-\alpha_{\rm inj}$ coordinate.}
\end{figure}

In the classical spectral ageing analysis the CI model includes the continuum
injection of relativistic particles having all the time the same power-law energy
distribution. In the presence of energy losses due to magnetic fields the observed spectrum,
being a sum of the emission from the various particle populations at different synchrotron
ages, is characterized by a curvature between its low-frequency and high-frequency ends. 
This curvature is formed by many spectra with different normalizations and break
frequencies. Implicit in this model is the assumption that the field strength is the same
in the region of injection as in the outflow region.
But the DYNAGE model is different because it accounts for the evolution of the magnetic
field strength in a different way than the original CIE model does that. Hence the
spectra calculated with it have different curvature.
Probably the actual spectra have some curvature in them as well. SYNAGE cannot cope
with such a curvature, thus fits a spectrum, single-broken between values of an
$\alpha_{\rm CI}$ and $\alpha_{\rm CI}+0.5$, to the data points. However, because
the observed spectrum is already affected by radiative losses below the fitted break,
$\alpha_{\rm CI}$ (and $\alpha_{\rm CIE}$) comes out rather steep. Contrary to that, 
DYNAGE can cope with the above curvature because, basing on a more realistic model of
dynamical expansion of the source (cocoon), it can better predict its 
spectrum. Therefore it can fit an aged spectrum with a flatter $\alpha_{\rm inj}$,
because the part of the spectrum already affected by radiative losses extends to much
lower frequencies than in the case of the classical approach. However, it should be
noted that the KDA model for the cocoon's emission is based on the approximation of the
synchrotron kernel radiation with a `delta'-function. This approximation could artificially
enhanced the spectral curvature, in the sense that it could result much more pronounced
than it really is.

Resuming, we argue that usually DYNAGE, finding a flatter $\alpha_{\rm inj}$,
is also less sensitive to its exact value. In other words, DYNAGE can afford radiative
effects at lower frequencies than SYNAGE. Radiative effects at lower
frequencies means older age, thus one can suspect that this effect is stronger at older
ages than at younger ages where the spectra tend to be `straighter', i.e. with only a
single break and less curvature.  But our, though very limited, statistics shows
something else. The difference $\alpha_{\rm CIE}-\alpha_{\rm inj}$ seems to depend on
redshift but not on the age. The correlation coefficient in the correlation between this
difference and the redshift is +0.63. Again, this effect can  support our thesis that
DYNAGE can better afford radiative losses at low frequencies, where an intrinsic spectral
curvature is shifted by the redshift, than the classical spectral ageing analysis.

(2) There is another factor having an influence on the difference between $t$ and $\tau$.
This is the axial ratio $R_{\rm T}$. Note that in the classical energy equipartition formula,
$B_{\rm eq}\propto (L/V_{\rm c})^{2/7}$, the cocoon's (lobe's) volume is completely
insensitive of a value of $R_{\rm T}$ what originate from the spherical geometry assumed
in the classical formula of Pacholczyk (1970). Oppositely,  because of the cylindrical
geometry assumed in KDA and DYNAGE, the cocoon's volume is strictly dependent of $R_{\rm T}$
which is $V_{\rm c}\propto R_{\rm T}^{-2}$ (its length $D$ is measured much more
precisely than the transversal base diameter). A variation of $R_{\rm T}$ results in a change
of the pressure ratio ${\cal P}_{\rm hc}$ which determines the lobe (cocoon) energy density.
This, in turn, specifies the magnetic field strength via the magnetic energy density (cf.
equations (3), (4), and (5)). The data in Tables 3 and 4 confirm that the smaller the differences 
$\alpha_{\rm CIE}-\alpha_{\rm inj}$ (and closer to unity the ratio between magnetic field
strength calculated with the formula of Miley (1980) and the field determined with equation
(5) in this paper, $B_{\rm DYN}$), the closer the dynamical and radiative ages.

\section{acknowledgements}

J.M. thanks Dr. Christian R. Kaiser for helpful discussions and constructive suggestions,
and the anonymous referee for the criticism
which allowed us to improve the paper. JM and MJ acknowledge the MNiSW funds for scientific
research in years 2005-2007 under contract No. 0425/PO3/2005/29.

\end{document}